\pdfoutput=1

\documentclass[11pt,twoside,a4paper,cmspaper,final,collab]{cms-tdr}
\def\svnVersion{46397M}\def\cmsCernNo{2010-092}\def\cmsCernDate{\today}\def\cmsMessage{Submitted to Physical Review Letters}

\begin{document}\cmsNoteHeader{QCD-10-016}

\hyphenation{had-ron-i-za-tion}
\hyphenation{cal-or-i-me-ter}
\hyphenation{de-vices}
\RCS$Revision: 46318 $
\RCS$HeadURL: svn+ssh://alverson@svn.cern.ch/reps/tdr2/papers/QCD-10-016/trunk/QCD-10-016.tex $
\RCS$Id: QCD-10-016.tex 46318 2011-03-17 19:50:48Z apana $

\newcommand{\CHI}{\ensuremath{\chi_\text{dijet}}\xspace}
\newcommand{\YSTAR}{\ensuremath{y^{\star}}\xspace}
\newcommand{\YBOOST}{\ensuremath{y_\text{boost}}\xspace}
\newcommand{\intlumi} {\ensuremath{36\ \textrm{pb}^{-1}}\xspace}
\newcommand{\mjj} {\ensuremath{M_{jj}}\xspace}
\newcommand{\ptrm}{$\text{p}_\text{T}$\xspace}
\newcommand{\ptit}{\ensuremath{p_{T}}\xspace}
\newcommand{\ptave}{\ensuremath{\langle p_{T}  \rangle}\xspace}
\newcommand{\ptavens}{\ensuremath{\langle p_{T}  \rangle}}

\cmsNoteHeader{QCD-10-016} % This is over-written in the CMS environment: useful as preprint no. for export versions
\title{\texorpdfstring{Measurement of Dijet Angular Distributions and Search for Quark Compositeness in pp Collisions at $\sqrt s~=~7$~TeV}{Measurement of Dijet Angular Distributions and Search for Quark Compositeness in pp Collisions at 7 TeV}}

\address[neu]{Northeastern University}
\address[fnal]{Fermilab}
\address[cern]{CERN}
\author[cern]{The CMS Collaboration}

\date{\today}

\abstract{Dijet angular distributions are measured over a wide
range of dijet invariant masses in pp collisions at $\sqrt s$ =
7~TeV, at the CERN LHC. The event sample, recorded with the CMS
detector, corresponds to an integrated luminosity of \intlumi. The
data are found to be in good agreement with the predictions of
perturbative QCD, and yield no evidence of quark compositeness.
With a modified frequentist approach, a lower limit on the contact
interaction scale for left-handed quarks of $\Lambda^{+} =
5.6$~TeV ($\Lambda^{-} = 6.7$~TeV) for destructive (constructive)
interference is obtained at the 95\% confidence level.}

\hypersetup{%
pdfauthor={CMS Collaboration},%
pdftitle={Measurement of Dijet Angular Distributions and Search for Quark Compositeness in pp Collisions at 7 TeV},%
pdfsubject={CMS},%
pdfkeywords={CMS, physics, software, computing}}

\maketitle %maketitle comes after all the front information has been supplied

In the standard model, point like parton-parton scatterings in
high energy proton-proton collisions can give rise to final states
with energetic jets. At large momentum transfers, events with at
least two energetic jets (dijets) may be used to confront the
predictions of perturbative Quantum Chromodynamics (pQCD) and to
search for signatures of new physics. In parton-parton scattering,
the angular distribution of the outgoing partons, $d
\hat{\sigma}/d \cos\theta^*$, is directly sensitive to the spin of
the exchanged particle, where $\hat{\sigma}$ is the parton-level
cross section and $\theta^{*}$ is the polar scattering angle in
the parton-parton center-of-mass (CM) frame. While QCD predicts a
noticeable deviation of the dijet angular distribution from
Rutherford scattering, at small CM scattering angles the angular
distribution is proportional to the Rutherford cross section, $
d\hat{\sigma} / d \cos\theta^* \sim 1 / (1 - \cos\theta^{*})^{2}$,
characteristic of spin-1 particle exchange. The dijet angular
distributions do not strongly depend on the details of the parton
distribution functions (PDFs), since the angular distributions for
the underlying processes, $qg \rightarrow qg$, $qq' \rightarrow
qq'$, and $gg \rightarrow gg$, are similar.

For the scattering of massless partons, which are assumed to be
collinear with the beam protons, the longitudinal boost of the
parton-parton CM frame with respect to the proton-proton CM frame,
$y_{\rm boost}$, and $\theta^{*}$ are obtained from the rapidities
$y_1$ and $y_2$ of the jets from the two scattered partons by
$y_{\rm boost} = {\frac{1}{2}}(y_1 + y_2)$ and $|\cos\theta^*| =
\tanh y^*$, where $y^{*} = {\frac{1}{2}}|y_1 -y_2|$ and where $\pm
y^{*}$ are the rapidities of the two jets in the parton-parton CM
frame. The rapidity is related to the jet energy $E$ and the
projection of the jet momentum on the beam axis $p_z$ by $y =
{\frac{1}{2}} \ln [(E + p_z) / (E - p_z)]$. The variable
$\chi_\text{dijet} = \exp(2y^{*})$ is used to measure the dijet
angular distribution, which for collinear massless-parton
scattering takes the form $\chi_\text{dijet} = (1 +
|\cos{\theta^*}|) / ( 1 - |\cos{\theta^*}|)$. This choice of
$\chi_\text{dijet}$, rather than $\theta^{*}$, is motivated by the
fact that $d\sigma_\text{dijet}/d\chi_\text{dijet}$ is flat for
Rutherford scattering. It also allows signatures of new physics
that might have a more isotropic angular distribution than QCD
(e.g. quark compositeness) to be more easily examined as they
would produce an excess at low values of $\chi_\text{dijet}$. The
quantity studied in this analysis is
 $(1/ \sigma_\text{dijet})(d\sigma_\text{dijet}/d\chi_\text{dijet})$, for
several ranges of the dijet invariant mass $M_{jj}$. Previous
searches for quark compositeness using the dijet angular
distribution or related observables in pp and p$\bar {\rm p}$
collisions have been reported at the Sp$\bar {\rm p}$S by the
UA1~\cite{bib_UA1} collaboration, at the Tevatron by the
D0~\cite{bib_d0-prl-chi,bib_d0-prd-chi} and CDF
~\cite{bib_cdf-prl-chi} collaborations, and at the Large Hadron
Collider (LHC) by the ATLAS~\cite{bib_atlas-chi} collaboration.
The CMS collaboration has also published a search on quark
compositeness with a smaller data sample using the dijet
centrality ratio~\cite{bib_cms-dijets}. In this Letter, we present
the first measurement of dijet angular distributions from CMS in
pp collisions at $\sqrt{s} = 7$~TeV.

The central feature of the CMS apparatus is a superconducting
solenoid, of 6~m internal diameter, providing an axial field of
3.8 T.  Within the field volume are the silicon pixel and silicon
strip tracker, the electromagnetic calorimeter (ECAL) and the
hadron calorimeter (HCAL). The ECAL is made up of lead-tungstate
crystals, while the HCAL is made of layers of plates of brass and
plastic scintillator. These calorimeters provide coverage in
pseudorapidity up to $|\eta| \leq 3$, where $\eta = - \ln
\tan(\theta/2)$ and $\theta$ is the polar angle relative to the
counterclockwise proton beam direction. An iron/quartz-fiber
\v{C}erenkov hadron calorimeter (HF) covers pseudorapidities $3 <
|\eta| < 5$. In addition, a preshower detector made of silicon
sensor planes and lead absorbers is located in front of the ECAL
at $1.653 < |\eta| < 2.6$. The calorimeter cells are grouped in
projective towers of granularity in pseudorapidity and azimuthal
angle of $0.087 \times 0.087$ at central pseudorapidities, with
coarser granularity at forward pseudorapidities. Muons are
measured in gas-ionization detectors embedded in the steel
magnetic field return yoke. A detailed description of the CMS
detector can be found elsewhere~\cite{bib_CMS}.

Events were collected online with a two-tiered trigger system:
Level-1 (L1) and the High Level Trigger (HLT). For this study,
events were selected with five inclusive single-jet triggers, with
the following jet transverse momentum \ptit thresholds at L1
(HLT): 20~GeV (30~GeV), 30~GeV (50~GeV), 40~GeV (70~GeV), 60~GeV
(100~GeV), and 60~GeV (140~GeV). The jets at L1 and HLT were
reconstructed using energies measured by the ECAL, HCAL, and HF,
and were not corrected for the jet energy response of the
calorimeters. All except the highest-threshold jet trigger were
prescaled as the LHC instantaneous luminosity increased during the
course of data taking. In each case, the trigger efficiency was
measured as a function of dijet invariant mass \mjj using events
selected by a lower-threshold trigger. For the analysis, \mjj- and
\CHI-regions were chosen such that the trigger efficiencies
exceeded $99\%$.

Jets were reconstructed offline from energies measured in the
calorimeter towers using the anti-$k_T$
clustering-algorithm~\cite{bib_antikt} with a distance parameter
$R = 0.5$. Spurious jets from noise and non-collision backgrounds
were eliminated by loose quality criteria on the jet
properties~\cite{bib_jetid}. The jet four-momenta were corrected
for the non-linear response of the calorimeters~\cite{bib_newjec}.
The performance of the CMS detector with respect to jet
reconstruction is described in detail
elsewhere~\cite{CMS-PAS-JME-10-003}.

Events were required to have a primary vertex reconstructed within
24 cm of the detector center along the beam
line~\cite{bib_vertexpas}. Events having at least two jets were
selected and the two highest-\ptit jets were used to measure the
dijet angular distributions for different ranges in $M_{jj}$. We
required $\chi_\text{dijet}<16$ and $|y_{\rm boost}|<1.11$, thus
restricting the rapidities $y_1$ and $y_2$ of the two
highest-\ptit jets to be less than 2.5. Nine analysis ranges were
defined with the boundaries $ 0.25 < M_{jj} < 0.35$~TeV, $0.35<
M_{jj} < 0.5$~TeV, $0.5< M_{jj} < 0.65$~TeV, $0.65< M_{jj} <
0.85$~TeV, $0.85< M_{jj} < 1.1$~TeV, $1.1< M_{jj} < 1.4$~TeV,
$1.4< M_{jj} < 1.8$~TeV, $1.8< M_{jj} < 2.2$~TeV, and $M_{jj}>
2.2$~TeV. The data correspond to integrated luminosities of 0.4,
3.5, 9.2, and 19.8~pb$^{-1}$ for the lowest four $M_{jj}$ ranges
and $36$ pb$^{-1}$ for the remaining ones. The uncertainty on the
integrated luminosity has been estimated to be
11\%~\cite{bib_lumi}.

The dijet angular distributions are corrected for migration
effects in $\chi_\text{dijet}$ and  $M_{jj}$ due to the finite jet
energy and position resolutions of the detector. The correction
factors were determined using two independent Monte Carlo (MC)
samples: {\sc pythia} 6.422~\cite{bib_pythia} with tune {\sc
D6T}~\cite{bib_pythiatunes} and {\sc herwig++}
2.4.2~\cite{bib_herwig}.  The four-momentum, rapidity, and
azimuthal angle of each generated jet were smeared to reproduce
the measured resolutions. The ratio of the two dijet angular
distributions (the generated distribution and the smeared one)
determined the unfolding correction factors for a given MC sample
and for each $M_{jj}$ range. The average of the correction factors
for each $M_{jj}$ range from the two MC samples formed the final
unfolding correction applied to the data. The correction factors
change the normalized dijet angular distributions for all $M_{jj}$
ranges by less than 3\%. For each $M_{jj}$ range, the systematic
uncertainty associated with each correction factor was set at 50\%
of its value. This approach covers the variations of the unfolding
correction factors determined from {\sc herwig++} and different
{\sc pythia} tunes ({\sc D6T} and {\sc Z2}~\cite{bib_tunez2}) that
vary on their modelling of the jet kinematic distributions. The
use of a parameterized model to simulate the finite jet \ptit and
position resolutions of the detector, to determine the unfolding
correction factors, resulted in a systematic uncertainty.  This
was estimated to be less than 1\% for all $M_{jj}$ ranges and was
added in quadrature to the unfolding uncertainties.

The normalized dijet angular distributions are relatively
insensitive to many systematic effects, in particular they show
little dependence on the overall jet energy scale. However, since
$\chi_\text{dijet}$ depends on $y^{*}$, they are sensitive to the
rapidity dependence of the jet energy calibration. Typical values
for the jet energy scale uncertainties for the considered phase
space in the variables of jet \ptit and $\eta$ covered in this
analysis are between 3\% and 4\%~\cite{bib_newjec}. The
uncertainty on the $\chi_\text{dijet}$ distributions due to the
jet energy calibration uncertainties was found to be less than
2.5\%. The uncertainty on the dijet angular distributions from the
jet \ptit resolution uncertainty, estimated to be
10\%~\cite{CMS-PAS-JME-10-003}, was found to be less than 1\%. The
total systematic uncertainty on the $\chi_\text{dijet}$
distributions, calculated as the quadratic sum of the
contributions due to the uncertainties in the jet energy
calibration, the jet \ptit resolution, and the unfolding
correction, is less than 3\% for all $M_{jj}$ ranges.

\begin{figure}[htb]
\centering
\includegraphics[width=\linewidth]{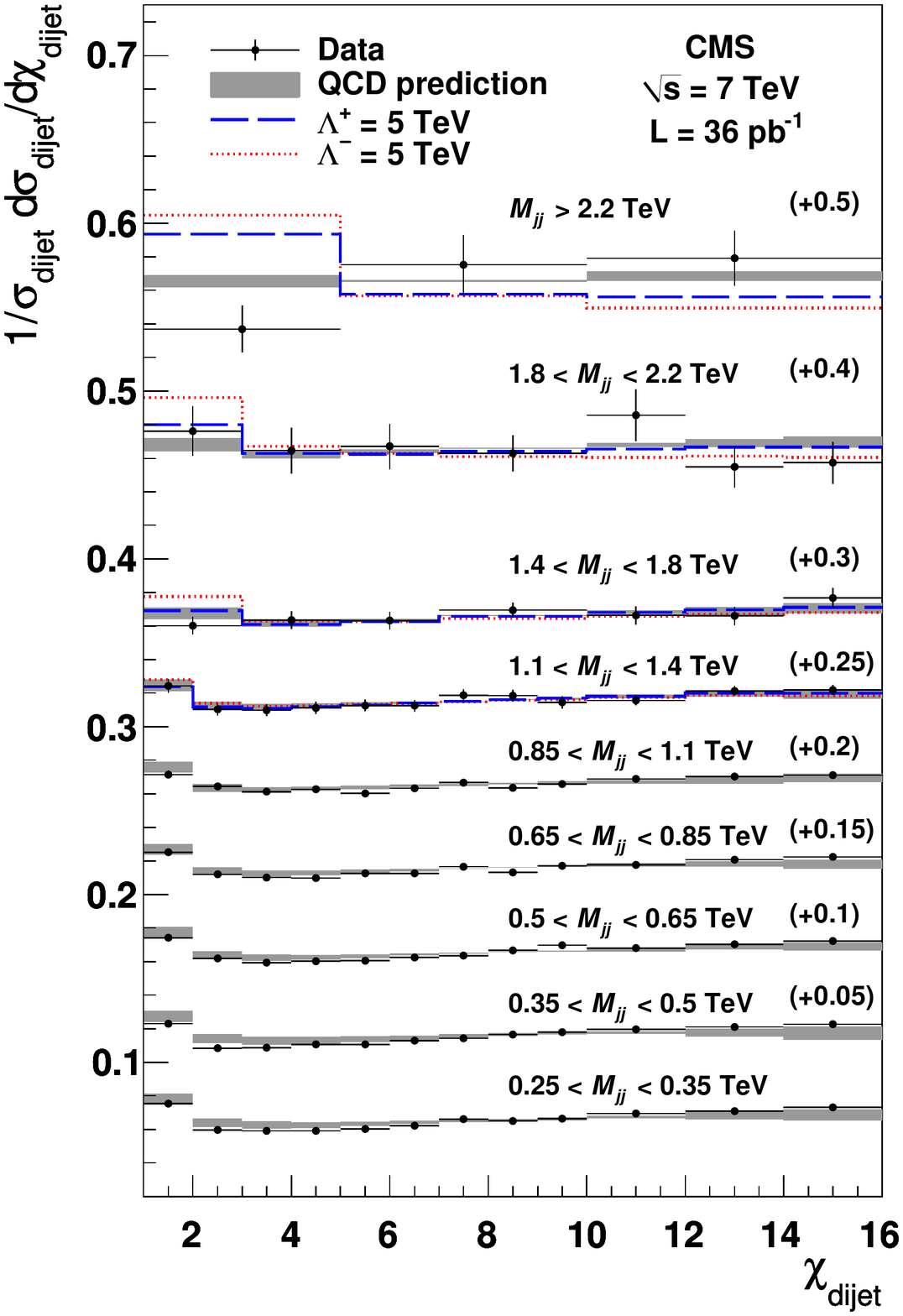}
\caption{\label{fig:chi}Normalized dijet angular distributions in
several $M_{jj}$ ranges, shifted vertically by the additive
amounts given in parentheses in the figure for clarity. The data
points include statistical and systematic uncertainties. The
results are compared with the predictions of pQCD at NLO (shaded
band) and with the predictions including a contact interaction
term of compositeness scale $\Lambda^{+}=5$~TeV (dashed histogram)
and $\Lambda^{-}=5$~TeV (dotted histogram). The shaded band shows
the effect on the NLO pQCD predictions due to $\mu_r$ and $\mu_f$
scale variations and PDF uncertainties, as well as the
uncertainties from the non-perturbative corrections added in
quadrature.}
\end{figure}

The corrected differential dijet angular distributions for
different $M_{jj}$ ranges, normalized to their respective
integrals, are shown in Fig.~\ref{fig:chi}. The data are compared
to pQCD predictions at next-to-leading order (NLO) calculated with
{\sc nlojet++}~\cite{bib_nlojet} in the {\sc
fastnlo}~\cite{fastnlo} framework. The calculations were performed
with the CTEQ6.6 PDFs~\cite{cteq}. The factorization ($\mu_{f}$)
and renormalization ($\mu_{r}$) scales were set to \ptave, the
average dijet \ptit. Non-perturbative corrections due to
hadronization and multiple parton interactions, determined using
the average correction from {\sc pythia} ({\sc D6T} tune) and {\sc
herwig++}, were applied to the prediction. The uncertainties on
the pQCD predictions, indicated by the shaded band in
Fig.~\ref{fig:chi}, are less than 6\% (9\%) at low (high)
$M_{jj}$. These uncertainties include contributions due to scale
variations and PDF uncertainties, as well as the uncertainties
from the non-perturbative corrections. The uncertainty due to the
choice of $\mu_{f}$ and $\mu_{r}$ scales was evaluated by varying
the default choice of scales in the following six combinations:
$(\mu_{f}$, $\mu_{r})$ = $($\ptave/2, \ptave/2$)$, $($\ptave/2,
\ptavens$)$, $($\ptave, \ptave/2$)$, $($2\ptave, 2\ptavens$)$,
$($2\ptave, \ptavens$)$ and $($\ptave, 2\ptavens$)$. These scale
variations modify the predictions of the normalized
$\chi_\text{dijet}$ distributions by less than 5\% (9\%) at low
(high) $M_{jj}$. The uncertainty due to the choice of PDFs was
determined from the 22 CTEQ6.6 uncertainty eigenvectors using the
procedure described in Ref.~\cite{cteq}, and was found to be less
than 0.5\% for all $M_{jj}$ ranges. Half the difference between
the non-perturbative corrections from {\sc pythia} and {\sc
herwig++} was taken as the systematic uncertainty, and was found
to be less than 4\% (0.1\%) at low (high) $M_{jj}$. Overall there
is good agreement between the measured dijet angular distributions
and the theoretical predictions for all $M_{jj}$ ranges.

The measured dijet angular distributions can be used to set limits
on quark compositeness represented by a four-fermion contact
interaction term in addition to the QCD Lagrangian. The value of
the mass scale $\Lambda$ characterizes the strengths of the quark
substructure binding interactions and the physical size of the
composite states. A color- and isospin-singlet contact interaction
(CI) of left-handed quarks gives rise to an effective Lagrangian
term: ${\cal{L}}_{qq}=\eta_0
(2\pi/\Lambda^2)(\overline{q}_{L}\gamma^{\mu}q_{L})(\overline{q}_{L}\gamma_{\mu}q_{L})$~\cite{CI-1,
CI-2}, where $\eta_0=+1$ corresponds to destructive interference
between the QCD and the new physics term, and $\eta_0=-1$ to
constructive interference. We investigate a model in which all
quarks are considered composite as implemented in the {\sc pythia}
event generator.

The contributions of the CI term in {\sc pythia} are calculated to
leading order (LO), whereas the QCD predictions for the dijet
angular distributions are known up to NLO. In order to account for
this difference in the QCD plus CI prediction, the cross-section
difference $\sigma^{\text{QCD}}_{\text{NLO}} -
\sigma^{\text{QCD}}_{\text{LO}}$ was added to the LO QCD+CI
prediction in each $M_{jj}$ and $\chi_\text{dijet}$ bin. With this
procedure, we obtain a QCD+CI prediction where the QCD terms are
corrected to NLO while the CI terms are calculated at LO.
Non-perturbative corrections due to hadronization and multiple
parton interactions were also applied to the prediction. The
prediction for QCD+CI at the scale of $\Lambda^{+}= 5$~TeV
($\eta_0=+1$) and $\Lambda^{-}= 5$~TeV ($\eta_0=-1$) are shown in
Fig.~\ref{fig:chi}, for the four highest \mjj ranges.

\begin{figure}[htb]
\centering
\includegraphics[width=\linewidth]{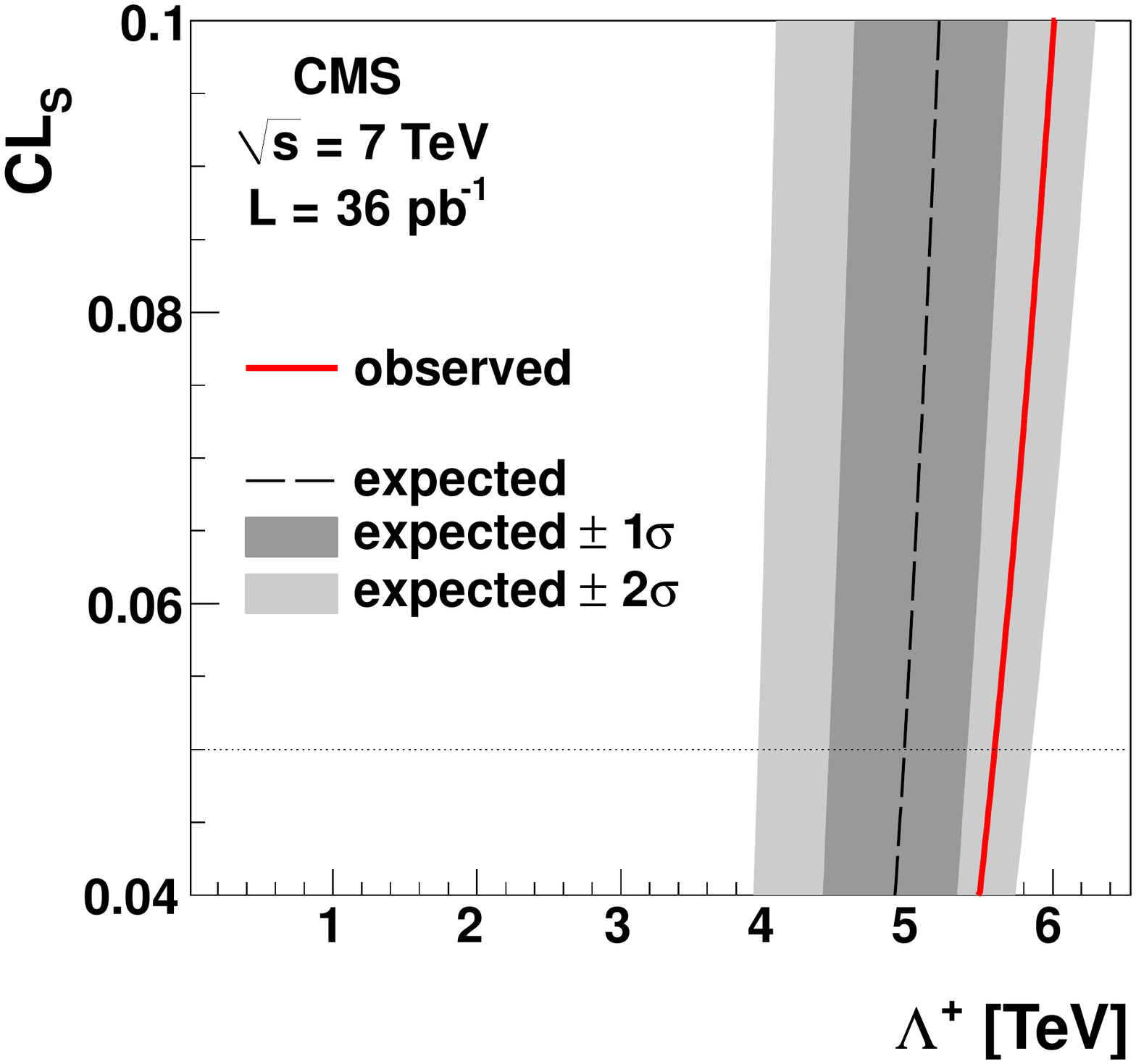}
\caption{\label{fig:limits}Observed ${\rm CL}_{\rm s}$ (solid
line) and expected ${\rm CL}_{\rm s}$ (dashed line) with one (two)
standard deviation(s) indicated by the dark (light) band as a
function of the contact interaction scale $\Lambda^{+}$. The 95\%
confidence level limits on $\Lambda^{+}$ are extracted from the
intersections of the observed and expected ${\rm CL}_{\rm s}$
lines with the horizontal line at ${\rm CL}_{\rm s}$=0.05.}
\end{figure}

We perform a statistical test discriminating between the QCD-only
hypothesis and the QCD+CI hypothesis as a function of the scale
$\Lambda$ based on the log-likelihood-ratio $Q = -2 \ln
(\frac{L_\text{QCD+CI}}{L_\text{QCD}})$. The likelihood functions
$L_\text{QCD+CI}$ and $L_\text{QCD}$ are modelled as a product of
Poisson likelihood functions for each bin in $\chi_\text{dijet}$
and \mjj in the four highest \mjj ranges. The prediction for each
\mjj range is normalized to the number of data events in that
range. The p-values, $P_\text{QCD+CI}(Q \geq Q_\text{obs})$ and
$P_\text{QCD}(Q \leq Q_\text{obs})$, are obtained from ensembles
of pseudo-experiments. A modified frequentist
approach~\cite{junk,read, read_physg} based on the quantity

\begin{equation*}
{\rm CL}_{\rm s} = \frac{P_\text{QCD+CI}(Q \geq Q_\text{obs})}
{1-P_\text{QCD}(Q \leq Q_\text{obs})}
\end{equation*}

is used to set limits on $\Lambda$. This approach is more
conservative than a pure frequentist approach (Neyman
construction) and prevents an exclusion claim when the data may
have little sensitivity to new physics~\cite{bib_PDG}. Systematic
uncertainties were introduced via Bayesian
integration~\cite{cousinshighland} by varying them as nuisance
parameters in the ensembles of pseudo-experiments according to a
Gaussian distribution convoluted with the shape variation induced
to the $\chi_\text{dijet}$ distributions. We consider the QCD+CI
model to be excluded at the 95\% confidence level if CL$_{\rm s} <
0.05$. Figure~\ref{fig:limits} shows the observed and expected
CL$_{\rm s}$ as a function of the CI scale $\Lambda^{+}$. From
this we determine the lower limit on $\Lambda^{+}$ to be 5.6~TeV.
The observed limit agrees within 1.4 standard deviations with the
expected limit of $5.0$~TeV, which was evaluated at the median of
the test statistics distribution of the QCD model. The observed
limit is slightly higher than the expected one because, for the
range $M_{jj} > 2.2$~TeV, the measured dijet angular distribution
at low $\chi_\text{dijet}$ is lower than, although statistically
compatible with, the QCD prediction. The limit for the CI scale
was also extracted using an alternate procedure in which the data
were not corrected for detector effects and instead the MC
predictions were resolution-smeared. The limit obtained was found
to agree with the quoted one within 0.4\%.  The corresponding
observed and expected limits on $\Lambda^{-}$ are 6.7~TeV and
5.8~TeV, respectively.

Shortly before the completion of this Letter, an exact NLO
calculation of QCD effects to quark compositeness became
available~\cite{gao_nlo}. This calculation indicates that the
limit on $\Lambda^{+}$ obtained in the present analysis, which
only takes into account the LO prediction for the contribution of
the contact interaction, might be overestimated by up to 10\%
compared to the value obtained if the NLO calculation were used.

In summary, CMS has measured the dijet angular distributions over
a wide range of dijet invariant masses. The $\chi_\text{dijet}$
distributions are found to be in good agreement with NLO pQCD
predictions, and are used to exclude a range of a color- and
isospin-singlet contact interaction scale $\Lambda$ for a
left-handed quark compositeness model. With a modified frequentist
approach, a lower limit on the contact interaction scale of
$\Lambda^{+} = 5.6$~TeV ($\Lambda^{-} = 6.7$~TeV) for destructive
(constructive) interference at the 95\% confidence level is
obtained, which may be compared with a limit of $5.0$~TeV
(5.8~TeV) expected for the number of events recorded. These are
the most stringent limits on the contact interaction scale of
left-handed quarks to date.

We wish to congratulate our colleagues in the CERN accelerator
departments for the excellent performance of the LHC machine. We
thank the technical and administrative staff at CERN and other CMS
institutes, and acknowledge support from: FMSR (Austria); FNRS and
FWO (Belgium); CNPq, CAPES, FAPERJ, and FAPESP (Brazil); MES
(Bulgaria); CERN; CAS, MoST, and NSFC (China); COLCIENCIAS
(Colombia); MSES (Croatia); RPF (Cyprus); Academy of Sciences and
NICPB (Estonia); Academy of Finland, ME, and HIP (Finland); CEA
and CNRS/IN2P3 (France); BMBF, DFG, and HGF (Germany); GSRT
(Greece); OTKA and NKTH (Hungary); DAE and DST (India); IPM
(Iran); SFI (Ireland); INFN (Italy); NRF and WCU (Korea); LAS
(Lithuania); CINVESTAV, CONACYT, SEP, and UASLP-FAI (Mexico); PAEC
(Pakistan); SCSR (Poland); FCT (Portugal); JINR (Armenia, Belarus,
Georgia, Ukraine, Uzbekistan); MST and MAE (Russia); MSTD
(Serbia); MICINN and CPAN (Spain); Swiss Funding Agencies
(Switzerland); NSC (Taipei); TUBITAK and TAEK (Turkey); STFC
(United Kingdom); DOE and NSF (USA).

\bibliography{auto_generated}   % will be created by the tdr script.

\begin{thebibliography}{10}
\bibitem{bib_tunez2}
The {\sc pythia6} {\sc Z2} tune is identical to the {\sc Z1} tune described
  in~\cite{bib_pythiatunes} except that {\sc Z2} uses the CTEQ6L PDF while {\sc
  Z1} uses CTEQ5L.
\end{thebibliography}
\cleardoublepage\appendix\section{The CMS Collaboration \label{app:collab}}\begin{sloppypar}\hyphenpenalty=5000\widowpenalty=500\clubpenalty=5000\textbf{Yerevan Physics Institute,  Yerevan,  Armenia}\\*[0pt]
V.~Khachatryan, A.M.~Sirunyan, A.~Tumasyan
\vskip\cmsinstskip
\textbf{Institut f\"{u}r Hochenergiephysik der OeAW,  Wien,  Austria}\\*[0pt]
W.~Adam, T.~Bergauer, M.~Dragicevic, J.~Er\"{o}, C.~Fabjan, M.~Friedl, R.~Fr\"{u}hwirth, V.M.~Ghete, J.~Hammer\cmsAuthorMark{1}, S.~H\"{a}nsel, C.~Hartl, M.~Hoch, N.~H\"{o}rmann, J.~Hrubec, M.~Jeitler, G.~Kasieczka, W.~Kiesenhofer, M.~Krammer, D.~Liko, I.~Mikulec, M.~Pernicka, H.~Rohringer, R.~Sch\"{o}fbeck, J.~Strauss, A.~Taurok, F.~Teischinger, P.~Wagner, W.~Waltenberger, G.~Walzel, E.~Widl, C.-E.~Wulz
% ...
\vskip\cmsinstskip
% ...
1:~~Also at CERN, European Organization for Nuclear Research, Geneva, Switzerland\\
% ...

\end{sloppypar}
\end{document}